\documentclass{aa}  

\usepackage{graphicx}
\usepackage{txfonts}
\usepackage{booktabs}
\usepackage{titlesec}
\usepackage[dvipsnames]{xcolor}
\usepackage{subfigure}
\usepackage{pdflscape}
\usepackage{ulem}
\usepackage{placeins}

\usepackage{color}
\usepackage{natbib,twoopt}
\usepackage[hyphenbreaks]{breakurl}
\usepackage[breaklinks]{hyperref}      
\bibpunct{(}{)}{;}{a}{}{,}             
\definecolor{cobalt}{rgb}{0.06, 0.2, 0.65}
\hypersetup{
  colorlinks,
  citecolor=blue,
  linkcolor=blue,
  urlcolor=blue,
}
\makeatletter
  \newcommandtwoopt{\citeads}[3][][]{\href{http://adsabs.harvard.edu/abs/#3}%
    {\def\hyper@linkstart##1##2{}%
     \let\hyper@linkend\@empty\citealp[#1][#2]{#3}}}
  \newcommandtwoopt{\citepads}[3][][]{\href{http://adsabs.harvard.edu/abs/#3}%
    {\def\hyper@linkstart##1##2{}%
     \let\hyper@linkend\@empty\citep[#1][#2]{#3}}}
  \newcommandtwoopt{\citetads}[3][][]{\href{http://adsabs.harvard.edu/abs/#3}%
    {\def\hyper@linkstart##1##2{}%
     \let\hyper@linkend\@empty\citet[#1][#2]{#3}}}
  \newcommandtwoopt{\citeyearads}[3][][]%
    {\href{http://adsabs.harvard.edu/abs/#3}
    {\def\hyper@linkstart##1##2{}%
     \let\hyper@linkend\@empty\citeyear[#1][#2]{#3}}}
\makeatother

\titleformat{\paragraph}
{\normalfont\normalsize}{\theparagraph}{1em}{}
\titlespacing*{\paragraph}
{0pt}{3.25ex plus 1ex minus .2ex}{1.5ex plus .2ex}

\newcommand{\bse}{\textsc{bse}\xspace}

\newcommand{\mocca}{\textsc{mocca}\xspace}

\newcommand{\fewbody}{\textsc{fewbody}\xspace}

\newcommand{\fg}{$\rm 1P$\xspace}
\newcommand{\sg}{$\rm 2P$\xspace}
\newcommand{\mpop}{$\rm MPOP$\xspace}
\titlerunning{MPOP with MOCCA: Spatial overconcentration of 1P RGB stars}
\authorrunning{M. Giersz et al.}

\begin{document}

\title{Multiple stellar populations in MOCCA globular cluster models: Transient spatial overconcentration of pristine red giant stars driven by strong dynamical encounters}

   \author{
    M. Giersz$^{1}$,  A. Askar$^{1}$,
     A Hypki$^{1,2}$, J. Hong $^{3}$, G. Wiktorowicz$^{1}$, L. Hellström$^{1}$
    }

   \institute{Nicolaus Copernicus Astronomical Center, Polish Academy of Sciences, ul. Bartycka 18, PL-00-716 Warsaw, Poland
   \\
              \email{mig@camk.edu.pl; askar@camk.edu.pl}
         \and
             Faculty of Mathematics and Computer Science, A. Mickiewicz University, Uniwersytetu Pozna\'nskiego 4, 61-614 Pozna\'n, Poland 
         \and
            Korea Astronomy and Space Science Institute, Daejeon 34055, Republic 
            of Korea         
             }

   \date{Accepted XXX. Received YYY; in original form ZZZ}

\abstract{Recent findings show that, in some Milky Way globular clusters (GCs), pristine red giant branch (RGB) stars are more centrally concentrated than enriched ones. This contradicts most multiple-population formation scenarios, which predict that the enriched population (\sg) should initially be more concentrated than the pristine population (\fg). 
We analyzed a MOCCA GC model that exhibits a higher spatial concentration of \fg RGB stars than \sg RGB stars at 13 Gyr. The MOCCA models assume the asymptotic giant branch (AGB) scenario, in which \sg stars are initially more concentrated than \fg stars. Our results indicate that the observed spatial distributions of multiple populations, and possibly their kinematics, are significantly shaped by dynamical interactions. These interactions preferentially eject \sg RGB progenitors from the central regions, leading to a transient overconcentration of \fg RGB stars at late times. This effect is particularly relevant for GCs with present-day masses of a few $10^5 \ \rm{M_{\odot}}$, which have retained only about \rm{10–20}\% of their initial mass. Such clusters may appear dynamically young due to heating from a black hole subsystem, even if they have undergone significant mass loss and dynamical evolution. Additionally, the relatively small number of RGB stars in these clusters suggests that interpreting the spatial distributions of multiple populations solely from RGB stars may lead to biased conclusions about the overall distribution of \fg and \sg.

The apparent overconcentration of the \fg relative to the \sg is likely a transient effect driven by the preferential removal of \sg RGB progenitors via strong dynamical encounters. MOCCA models of multiple stellar populations based on the AGB scenario may explain anomalous features observed in some Galactic GCs, such as NGC 3201 and NGC 6101.}


   \maketitle

\section{Introduction}
\label{s:intro}

Globular clusters (GCs) were historically thought to host simple stellar populations with uniform chemical compositions. However, photometric and spectroscopic studies over the past few decades have revealed the presence of multiple stellar populations (\mpop), primarily characterized by internal variations in light-element abundances \citep[see][and references therein]{Bastian2018,Gratton2019}. These populations include a pristine first population (\fg), whose chemical composition resembles that of field stars with similar metallicities, and one or more enriched populations (\sg), which exhibit signatures of high-temperature hydrogen burning \citep{Milone2022}. The fraction of enriched \sg stars in Milky Way (MW) GCs strongly correlates with the present-day cluster mass, ranging from approximately 40\% in low-mass clusters to nearly 90\% in the most massive ones \citep[see][and references therein]{Milone2022}.

The formation of \mpop in GCs remains a topic of active debate, with several theoretical scenarios proposed to explain its origin. However, none of these scenarios fully accounts for all the observational features reported in the literature. \citet{Gierszetal2024} provide a concise overview of the most widely accepted formation mechanisms, particularly from the perspective of numerical simulations. Their study also suggests that the post-formation migration of GCs may play a crucial role in shaping the properties of \mpop.
Most formation scenarios, including those that attribute \mpop to pollution from asymptotic giant branch (AGB) stars, predict that \sg should be more centrally concentrated than \fg. Observations generally support this prediction, but exceptions exist, such as the Galactic GCs NGC 3201 and NGC 6101, where \fg red giant branch (RGB) stars are found to be more centrally concentrated than \sg RGB stars \citep{Leitingeretal2023}. However, in the case of NGC 3201, \citet{Mehta2024}, using RGB stars along with other evolved stars from the horizontal branch (HB), found no evidence of such overconcentration, as \fg and \sg appear to be well mixed.

In this short paper, we present results from a GC model simulated with the \mocca code, in which the \sg population initially has a much higher central concentration than the \fg. However, by 13 Gyr, the \fg RGB stars become more spatially overconcentrated than the \sg RGB stars. The overconcentration of \fg RGB stars relative to \sg RGB stars may provide valuable insights into their formation scenarios and long-term evolution. It could also serve as an important constraint on the initial conditions of GC formation. Therefore, understanding the mechanisms driving this phenomenon and identifying the clusters where it may occur are crucial. In this study, we focus on the transient nature of \fg RGB overconcentration, offering a general explanation for its formation and the types of GCs in which it may be observed. A more detailed investigation will be presented in a forthcoming paper.

The paper is structured as follows. Section~\ref{s:Method} provides a brief overview of the \mocca code.  Section~\ref{s:Results} presents the simulation results for models exhibiting \fg overconcentration, and Section~\ref{s:Discussion} interprets these findings and discusses their implications.

\section{Method}
\label{s:Method}

This study is based on simulations carried out with the MOCCA Monte Carlo code, which models the stellar and dynamical evolution of realistic GCs, including multiple populations and delayed \sg formation. For details of the code and physical assumptions, see Appendix \ref{s:Appendix0}.

The model presented in this paper has the following initial conditions: the number of objects, $ N_{\rm1P}$ = \rm{800000}, $N_{\rm2P}$ = \rm{300000}, and a \citet{Kroupa2001} initial mass function (IMF), with \fg stars sampled between $\rm 0.08 \ M_{\odot}$ and $\rm 150 \ M_{\odot}$, and \sg stars between $\rm 0.08 \ M_{\odot}$ and $\rm 20 \ M_{\odot}$. The metallicities are $\rm Z_{1P}$ = \rm{0.001} and $\rm Z_{2P}$ = \rm{0.001}, with binary fractions $fb_{\rm1P}$ = \rm{0.95} and $fb_{\rm2P}$ = \rm{0.95}\footnote{Initial binary parameters were distributed according to \citet{Bellonietal2017}.}. The Galactocentric distance is $R_g$ = \rm{1.0} kpc, and the ratio of half-mass radii is $R_{h_{2P}}/R_{h_{1P}}$ = \rm{0.1}. \fg and \sg are separately in virial equilibrium, with \fg being tidally filling (with $R_{h_{1P}}=9.03$ pc and $R_{t}/ R_{h_{1P}}= 3.66$). The initial \citet{King1966} model concentration parameter is $\rm W_{0} = 3$ for \fg and $7$ for \sg. Over \rm{1} Gyr following \rm{5.5} Gyr of evolution, the cluster migrates along its circular orbit from \rm{1} kpc to \rm{5} kpc, mimicking the capture of a GC during a dwarf galaxy merger with the MW.

\section{Results}
\label{s:Results}

\begin{figure}
\begin{center}
  \includegraphics[width=0.99\linewidth]{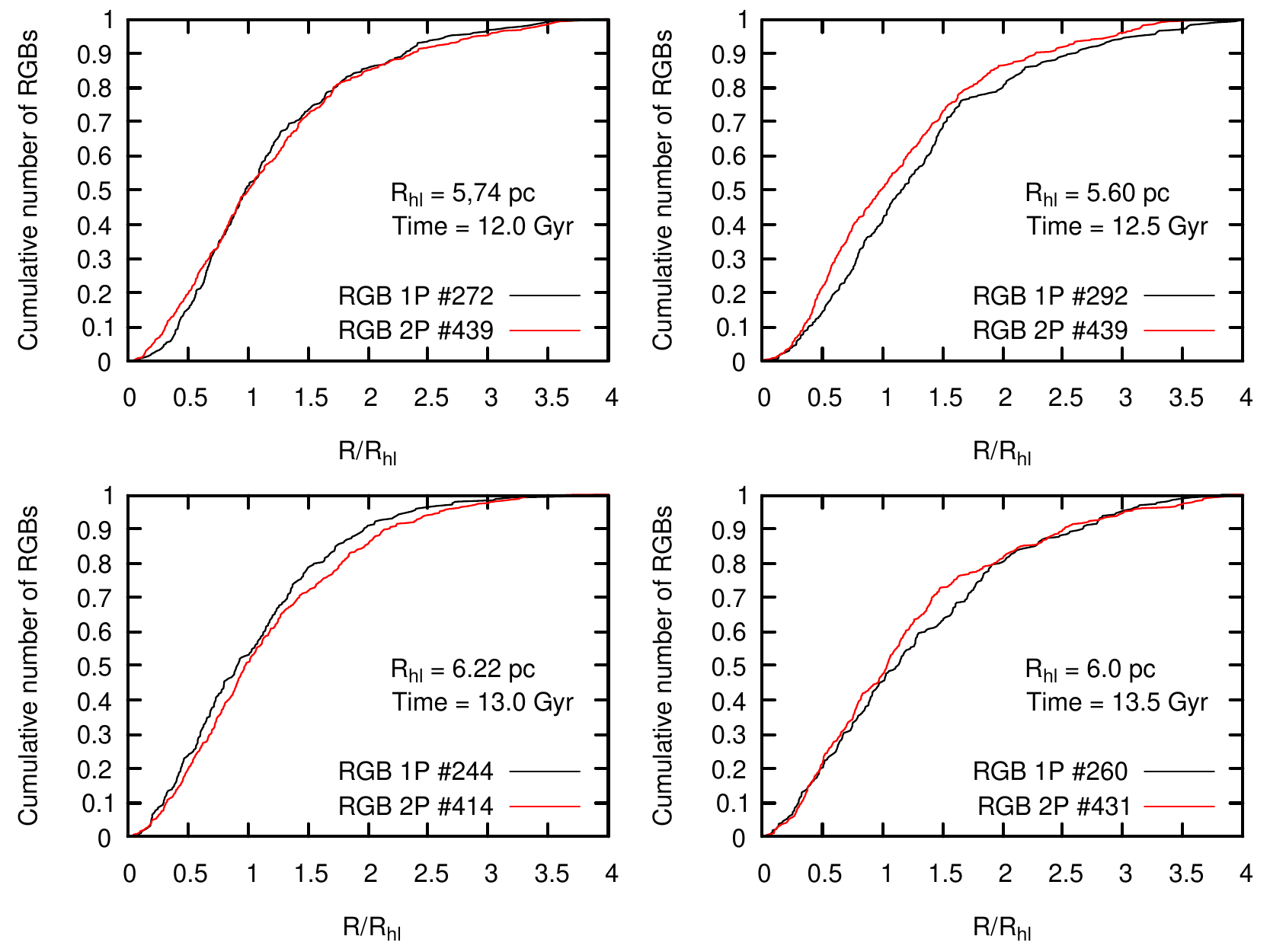}
  \caption{Cumulative number distributions of RGB stars for \fg (black line) and \sg (red line) as a function of distance scaled by $R_{hl}$ for selected projected (2D) snapshots from \rm{12} Gyr to \rm{13.5} Gyr. \fg is more centrally concentrated than \sg only in the snapshot at \rm{13} Gyr. The number distributions of both populations change significantly over time, and the overconcentration of \fg is a transient feature. The $R_{hl}$ values, snapshot times, and the number of RGB stars are provided in the insets of each panel.
  }
  \label{f:1:RGBtransient}
\end{center}
\end{figure}

Inspired by the work of \citet{Leitingeretal2023}, who conducted a homogeneous analysis of \mpop in 28 Galactic GCs using a combination of HST photometry and wide-field, ground-based photometry of RGB stars, we examine a puzzling trend that they identified. Their study found that in a few Galactic GCs, such as NGC 3201 and NGC 6101, \fg RGB stars are more concentrated within a few times the half-light radius than \sg RGB stars. This unexpected result is difficult to reconcile with current \mpop formation scenarios. Recently, \citet{Mehta2024} cast doubt on these findings, showing that in NGC 3201, when HB and AGB stars are included alongside RGB stars, both populations exhibit similar spatial distributions. However, \citet{Cadelano2024} confirmed the results of \citet{Leitingeretal2023} for NGC 3201, though they found that \sg RGB stars follow a bimodal distribution and are more concentrated within the half-light radius.

With these results in mind, we took a closer look at the \mpop simulations performed with the \mocca code, as is discussed in \citet{Gierszetal2024}. We identified several models in which \fg stars are more concentrated than \sg stars. Studies by \citet{Geisleretal1995,Massarietal2019,Chen2024} suggest that NGC 6101 and NGC 3201 likely formed ex situ, a scenario supported by the fact that both clusters follow retrograde orbits. Based on this, we selected a GC model with a capture time by the MW of \rm{5.5} Gyr and an orbital migration time of \rm{1} Gyr.
In addition, we tried to select a model with observational properties similar to those of NGC 3201, specifically: 
a cluster mass of \rm{1.45}$\times10^{5}$ (\rm{1.93}$\times10^{5}$) $M_{\odot}$, a core radius of $R_c$ = \rm{1.22} (\rm{1.66}) pc, an observational core radius of $R_{c,obs}$ = \rm{3.13} (\rm{1.85}) pc, a half-light radius of $R_{hl}$ = \rm{6.22} (\rm{5.17}) pc,
a half-mass radius of $R_h$ = \rm{8.71} (\rm{8.68}) pc, a tidal radius of $R_t$ = \rm{81.72} (\rm{82.34}) pc, and the ratio for RGB stars between the \fg and \sg populations of \rm{0.59} (\rm{0.46}). The values for the \mocca model were taken at \rm{13} Gyr, while the values in parentheses represent the observed properties of NGC 3201 from \citet{Baumgardt2018} and \citet{Miloneetal2017}.
The observational properties of the NGC 3201 are close to those of the model (except for $R_{c,obs}$ - see Appendix\,\ref{s:Appendix2} for a discussion), but this does not mean that the model represents the actual evolution of the GC. It only serves as a possible example of GC evolution, and we should not expect all cluster properties to be the same, in particular, the \fg concentration range. The model is used to illustrate the possible causes of the observed \fg and \sg properties in NGC 3201.

\begin{figure}
\begin{center}
  \includegraphics[width=0.99\linewidth]{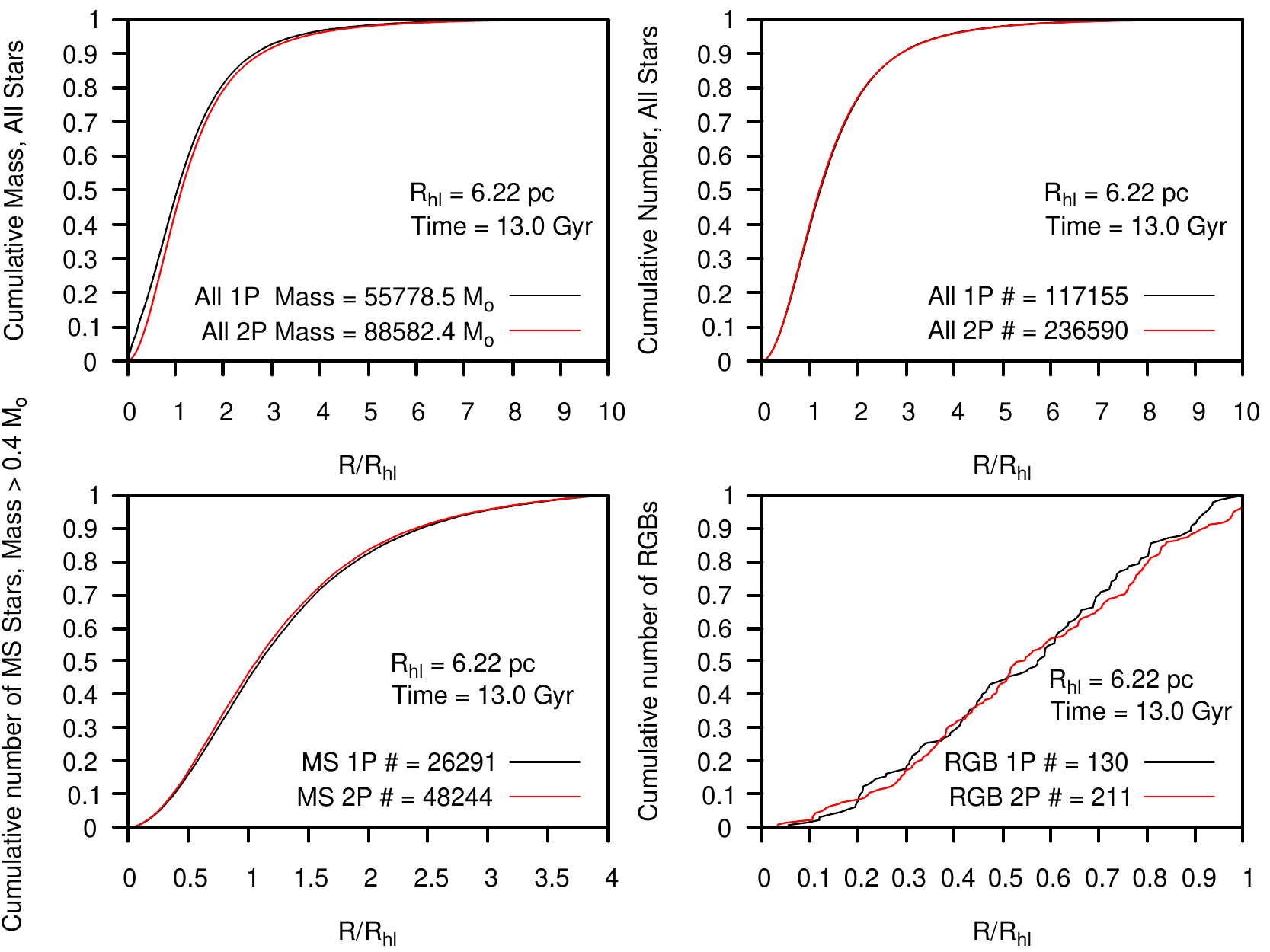}
  \caption{Top left panel: Cumulative mass distributions of all stars for \fg (black line) and \sg (red line) as a function of distance scaled by $R_{hl}$ in the projected (2D) snapshot at \rm{13} Gyr.
  Top right panel: Cumulative number distributions of all stars for \fg (black line) and \sg (red line) as a function of distance scaled by $R_{hl}$ in the snapshot at \rm{13} Gyr.
  Bottom left panel: Cumulative number distributions of MS stars with masses greater than \rm{0.4}$M_{\odot}$ for \fg (black line) and \sg (red line) as a function of distance scaled by $R_{hl}$ in the snapshot at \rm{13} Gyr.
  Bottom right panel: Cumulative number distributions of RGB stars for \fg (black line) and \sg (red line) as a function of distance scaled by $R_{hl}$ in the snapshot at \rm{13} Gyr, shown up to a distance of $R_{hl}$.
  }
  \label{f:2:CumulAllStars}
\end{center}
\end{figure}

Figure \ref{f:1:RGBtransient} shows the cumulative number distributions of RGB stars for \fg and \sg as a function of distance scaled by $R_{hl}$ at several selected snapshot times for a chosen \mocca model.
It clearly shows that \fg is more centrally concentrated than \sg only in the snapshot at \rm{13} Gyr. In other snapshots, both populations are well mixed, or \sg is more concentrated than \fg. The number distributions of both populations vary significantly over time, indicating that the overconcentration of \fg is a transient feature. The magnitude of \fg overconcentration at \rm{13} Gyr is not as pronounced as that observed for NGC 3201 by \citet{Leitingeretal2023} and \citet{Cadelano2024}. This is not unexpected, as the presented model is only a rough approximation of an actual GC. As is discussed in Section \ref{s:Discussion}, the extent of \fg overconcentration depends on the evolutionary details of RGB progenitors, the migration time, and the initial conditions of the cluster model.

Since the overconcentration of \fg relative to \sg appears to be a transient phenomenon, we examined whether this effect is also present when considering stellar types other than RGB stars. We selected the snapshot at \rm{13} Gyr, and as in Figure \ref{f:1:RGBtransient}, all distances are scaled by $R_{hl}$. 
Figure \ref{f:2:CumulAllStars} presents cumulative distributions for different stellar populations. We analyzed main sequence (MS) stars with masses greater than \rm{0.4} $M_{\odot}$ (a mass range typically observable in most GCs) as well as all stars, including compact remnants, belonging to \fg and \sg. The top left panel shows the cumulative mass distribution of all \fg and \sg stars, while the top right panel presents the cumulative number distributions of all stars for both populations. The bottom left panel displays the cumulative number distribution of MS stars with masses greater than \rm{0.4} $M_{\odot}$ for \fg and \sg.
Finally, the bottom right panel presents the cumulative number distributions of RGB stars for \fg and \sg up to a distance equal to $R_{hl}$. The overconcentration of \fg relative to \sg is also evident when analyzing the mass distribution of all stars in the cluster. While the overconcentration is subtle, it remains noticeable. For both populations, the half-mass radii were already mixed at 5 Gyr. However, when considering the number distribution of stars instead of the mass distribution, the two populations appear to be perfectly mixed. From the analysis of the model, it appears that stellar-mass black holes (BHs) are responsible for this effect. These BHs originate from \fg stars, as \sg stars have a maximum initial mass of only \rm{20}$M_{\odot}$, which significantly limits \sg BH formation due to the IMF. Over time, these BHs form a subsystem (BHS) at the cluster center. The influence of the BHS on the global evolution of the cluster is discussed in Section \ref{subsec:gc-model}. When considering only MS stars that are observable in distant GCs (e.g., those with masses greater than $\rm{0.4} \ M_{\odot}$), the overconcentration of \fg stars in the number distribution is not apparent, as both populations appear well mixed. This suggests that the type and number of selected stars play a crucial role in detecting the overconcentration of \fg relative to \sg.
\citet{Leitingeretal2023} and \citet{Cadelano2024} reported that in the central region of NGC 3201 (inside $R_{hl}$), \sg stars are more concentrated than \fg stars or are nearly mixed. A similar trend is observed in the bottom right panel of Figure \ref{f:2:CumulAllStars}. In the next section (Section \ref{s:Discussion}), we further analyze the properties of models that explain why \fg is more concentrated than \sg, why the overconcentration of \fg is observed only in RGB stars, and why this feature appears to be transient.

\section{Discussion}
\label{s:Discussion}

\subsection{Black hole subsystem and cluster migration}\label{subsec:gc-model}

By analyzing \mocca simulations with the \mpop implementation, as is described in \citet{Gierszetal2024}, and focusing on models in which \fg is more centrally concentrated than \sg, we draw the following conclusions. The overconcentration of \fg is observed in GC models that have lost 80–90\% of their initial mass, leading to present-day (11–13 Gyr) masses of approximately  
$\rm 10^5 \ M_{\odot}$. In all models that show 1P overconcentration, the feature is transient. Additionally, these clusters retain a BHS comprising up to a few hundred stellar-mass BHs at the present day. In clusters for which the present-day mass is only 10–20\% of the initial mass and the BHS accounts for approximately 5\% of the total cluster mass, the mass distribution of \fg becomes more centrally concentrated than that of \sg. All models with 1P overconcentration are characterized by a very wide range of tidal radii from 10-20 to about 80-90 pc. Only models with migration at a late time show properties similar to the ones observed in MW GCs.

The GC model described in this paper initially retains nearly 1000 stellar-mass BHs at around 50 Myr, originating from the evolution of massive \fg progenitors. At this time, the 10\% and 50\% Lagrangian radii for these BHs are approximately 4.8 pc and 10.8 pc, respectively, and begin to segregate toward the center, finally forming a BHS\ (see Appendix\,\ref{s:Appendix1} for a more detailed description of the BHS evolution). 
The BHS, consisting of both single and binary BHs, strongly influences the cluster’s dynamical evolution through interactions with other stars \citep{Mackey2008,BreenHeggie2013,Sippel2013,Morscher2015,ArcaSedda2018,Kremer2019,Weatherford2020}. Since the \sg population is initially more centrally concentrated in this model, \sg stars are more likely to undergo strong encounters with \fg BHs and other \sg stars. These interactions can scatter \sg stars onto wider orbits or eject them from the cluster entirely (see Section \ref{subsec:dyn-ejections}).  

For both NGC 3201 and NGC 6101, where \fg RGB stars appear more centrally concentrated than \sg RGB stars, several studies suggest that these GCs likely harbor a significant population of stellar-mass BHs \citep{Peuten2016,Askar2018-bhsubsys,Kremer2018,Vitral2022}. Furthermore, three stellar-mass BHs have been identified in NGC 3201 through radial velocity variations using MUSE spectroscopic observations \citep{Giesers2018,Giesers2019}. Therefore, it is possible that the possible presence of a BHS has played a role in shaping the spatial distribution of \fg and \sg RGB stars in these GCs. 

The GC NGC 3201 is likely an ex situ cluster that was accreted by the MW along with a dwarf galaxy in the Gaia-Enceladus--Sequoia event \citep{Massarietal2019}. It is characterized by large $R_h$ and $R_t$ radii. During the early stages of its evolution, it likely experienced a strong tidal field within its parent galaxy, and it can be assumed that it was initially tidally filling, with small $R_h$ and $R_t$ radii \citep{Renaudetal2017}. After being captured by the MW, the tidal field weakened considerably, causing the cluster to become tidally underfilled, allowing it to expand freely within its $R_t$ and significantly increase its half-mass relaxation time.   For these reasons, we selected a cluster model that initially forms near the galactic center with an initial mass of approximately $10^6 M_{\odot}$. Over a few gigayears, the cluster migrates to a larger Galactocentric distance (see Section \ref{s:Results} for details). This model not only reproduces similar observational properties to NGC 3201 but also yields a comparable ratio of \sg to \fg stars. We would like to strongly emphasize that GC migration time is a free parameter in simulations and plays an important role in creating conditions for achieving 1P overconcentration. At the same time, the magnitude of \fg overconcentration depends on many initial parameters such as the number ratio of \sg to \fg, the mass of the GC, the overconcentration of \sg relative to \fg, the galactocentric position of the GC, and the strength of the galactic tidal field. The capture time of NGC 3201 by the MW is about 10 Gyr \citep{Massarietal2019}. This is about twice as large as is assumed in the model. Since the strength of the 1P overconcentration depends on many initial parameters, we are confident that the appropriate selection of these parameters will result in a model of the NGC 3201 cluster that corresponds much better to observations. This will be the subject of the next paper.

Due to the cluster’s relatively low present-day mass, the number of RGB stars is relatively small, on the order of several hundred. As we show below, such a small number of RGB stars plays a crucial role in the overconcentration of \fg relative to \sg RGB stars.

\begin{figure}
\begin{center}
  \includegraphics[width=0.99\linewidth]{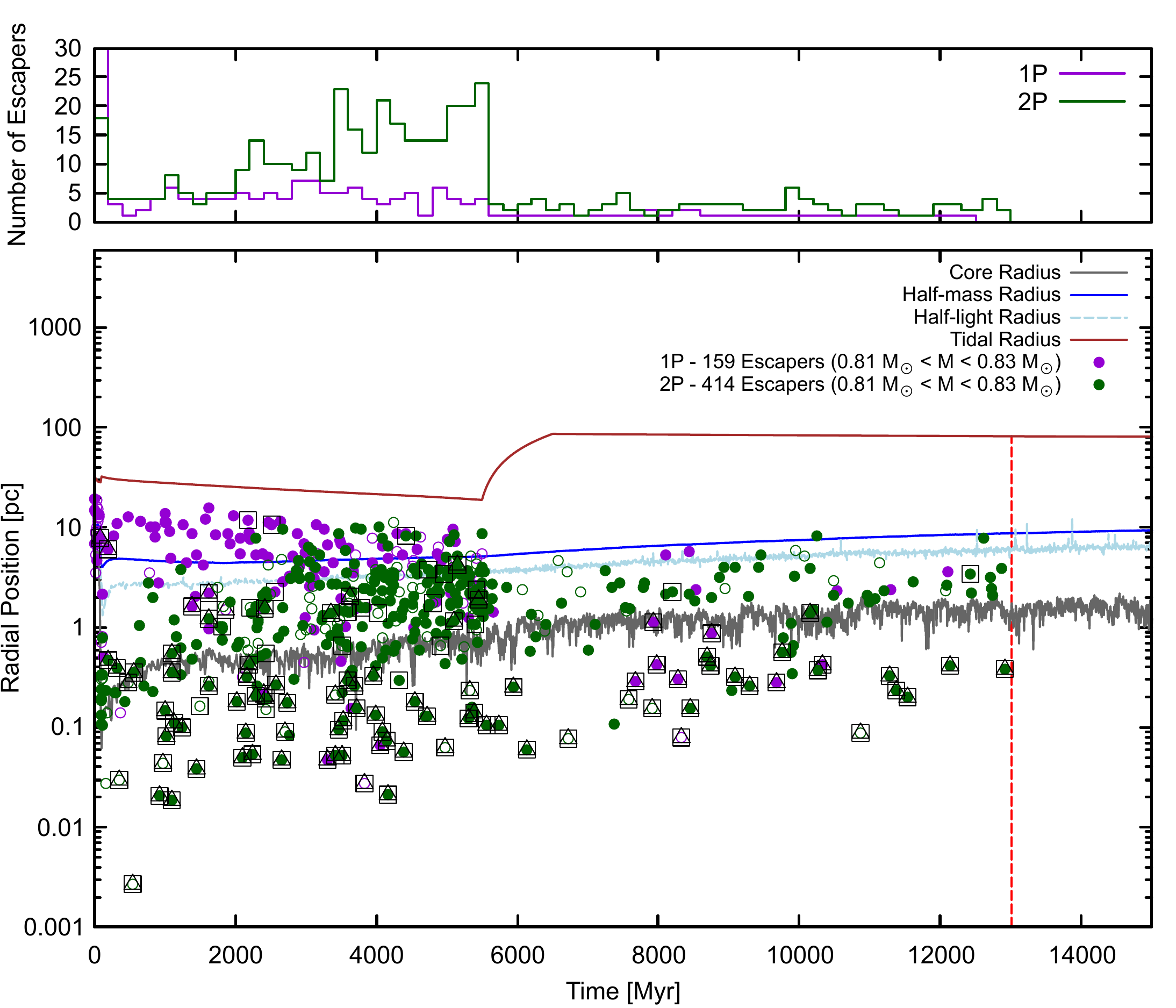}
\caption{The lower panel: Escape time versus radial position for dynamically ejected present-day RGB progenitor stars with initial masses between 0.81 and 0.83\,M$_\odot$. Each point represents a present-day RGB progenitor that escaped due to a strong three-body (binary–single) or four-body (binary–binary) encounter. The x axis shows the escape time, and the y axis indicates the radial position within the cluster at which the escape-causing interaction occurred. Superimposed lines trace the evolution of the core radius ($R_c$, grey), half-mass radius ($R_h$, dark blue), half-light radius ($R_{hl}$, light blue), and tidal radius ($R_t$, brown). Violet points represent 1P escapers, and green points represent 2P escapers; filled symbols denote single stars, and open symbols denote binaries. Triangles indicate stars that escaped directly due to an encounter involving at least one BH, while squares indicate stars that had at least one such BH encounter during their dynamical history. Of the 159 1P escapers, 20 (12.6\%) experienced BH interactions; among the 414 2P escapers, 96 (23.2\%) were involved in BH interactions. The vertical dashed red line marks the 13\,Gyr snapshot, where the 1P overconcentration is most evident. The top panel: Histograms of escape times for both populations, binned in 200\, Myr intervals, highlighting the higher escape rate of 2P RGB progenitors from strong encounters.}
  \label{f:3:time-vs-dyn-escapers}
\end{center}    
\end{figure}

\subsection{Preferential ejection of enriched RGB progenitors by strong dynamical encounters}\label{subsec:dyn-ejections}

To explain the spatial overconcentration of \fg RGB stars at \rm{13} Gyr, we first identified stars that evolve into RGBs at this epoch. These stars are near the turn-off mass, which is \rm{0.8154} $\rm M_{\odot}$ for $\rm Z=0.001$ and had a zero-age MS mass of approximately \rm{0.83} $\rm M_{\odot}$. Assuming that all stars in the initial model (or at $\sim$100 Myr, when \sg is fully added) with masses between \rm{0.81} and \rm{0.83} $\rm M_{\odot}$ will have evolved into RGB stars by \rm{13} Gyr, we defined this mass range as the present-day RGB progenitors and examined those that escaped before \rm{13} Gyr.

For \fg stars, we counted all escaping RGB progenitors within the \rm{0.81–0.83} $\rm M_{\odot}$ mass range, regardless of the cause of escape, and found that \rm{5048} \fg RGB progenitors had escaped the cluster. Similarly, for \sg stars, \rm{1167} RGB progenitors had escaped. This significant difference is expected, as \fg was initially tidally filling. However, when considering only escapers due to strong dynamical interactions (binary-single or binary-binary encounters), the numbers shift dramatically: \fg RGB progenitor escapers drop to \rm{159}, while \sg increases to \rm{414}. Thus, \sg RGB progenitors are \rm{2.6} times more likely to be dynamically ejected than \fg progenitors, which is consistent with their initial central concentration.

We find that \sg RGB progenitors are more frequently ejected from the inner regions of the cluster via strong dynamical encounters. As is shown in Fig.\,\ref{f:3:time-vs-dyn-escapers}, most \sg escapers are removed between \rm{2} and \rm{6} Gyr, primarily from within a few times the half-light radius ($R_{hl}$). After \rm{6} Gyr, following the cluster’s migration to a larger Galactocentric distance, the ejection rate declines but remains higher for \sg stars. The lower panel shows that 2P escapers are mostly launched from the inner regions, while the top histogram highlights their higher escape rate compared to \fg. Markers in the figure also reveal that a significant fraction of these ejections -- especially for 2P stars -- involve interactions with BHs (23.2\% for 2P vs. 12.6\% for 1P). Together, these results confirm that strong dynamical encounters, often involving BHs, preferentially remove \sg RGB progenitors, contributing to the overconcentration of \fg RGB stars at \rm{13} Gyr.

To quantify this effect, we defined the ratio:
\[
R = \frac{N_\text{RGB,in\_cluster}}{N_\text{RGB,progenitor\_dyn\_escapers}},
\]
where \( N_\text{RGB,in\_cluster} \) is the number of RGB stars present in the cluster at a given snapshot, and \( N_\text{RGB,progenitor\_dyn\_escapers} \) is the cumulative number of RGB progenitors lost to strong encounters by that time. For \sg at \rm{13} Gyr, we find \( R = 1.04 \) (\rm{430/414}), indicating that the number of remaining RGB stars is nearly equal to those lost through strong encounters. At earlier times, such as \rm{12.5} Gyr, this ratio is higher (\( R = 1.10 \)), as fewer progenitors have escaped relative to those still in the cluster. By \rm{13.5} Gyr, the ratio rises to \( R = 1.11 \), showing that the overconcentration of \fg relative to \sg RGB stars is most pronounced at \rm{13} Gyr, when \( R \) is closest to unity. This near-parity magnifies the impact of missing \sg RGB progenitors on the observed spatial distributions. At \rm{13} Gyr, the number of 2P RGB stars in the cluster (\rm{430}) is nearly matched by the number of progenitors ejected via strong encounters (\rm{414}). By contrast, at \rm{12.5} Gyr, the higher ratio (\( R = 1.1 \)) means more 2P RGB stars remain in the cluster, lessening this effect. By \rm{13.5} Gyr, \( R \) increases again to 1.11, diminishing the overconcentration. This increase occurs because lower-mass RGB progenitors, more numerous due to the IMF and still present in the cluster, begin evolving onto the RGB at this stage, thereby increasing the numerator in the ratio.

This analysis highlights the distinct dynamical evolution of \sg RGB progenitors, which are preferentially ejected through strong interactions, particularly from the cluster's inner regions. The interplay of small-number statistics and dynamical encounters thus plays a crucial role in shaping the observed RGB spatial distribution.

\subsection{Conclusions}\label{conclusion}

Our study suggests that the spatial distribution and, potentially, the kinematic properties of \mpop may depend on the type of stars observed. This effect is particularly relevant for GCs with present-day masses of a few $10^5 \ \rm{M_{\odot}}$, which have retained only \rm{10–20}\% of their initial mass. Despite the significant mass loss, such clusters may appear dynamically young due to BHS heating and migration to larger Galactocentric distances. When observations are limited to RGB stars, small-number statistics and dynamical interactions can distort their spatial distribution, leading to biased conclusions about the overall distribution of \mpop. The overconcentration of \fg RGB stars relative to \sg RGB stars appears to be a transient effect.  

Confirming the findings of \citet{Leitingeretal2023} and \citet{Cadelano2024}, that in some MW GCs, \fg RGB stars are more centrally concentrated than \sg RGB stars, requires observations of MS stars. Our \mocca simulations show no \fg overconcentration relative to \sg among MS stars. Verifying this observationally would be a crucial step toward understanding \mpop properties and supporting the AGB scenario, and possibly other scenarios, of \mpop formation. Additionally, our results align with the kinematic analysis by \citet{Cadelano2024}, which suggests that \sg stars initially formed more centrally concentrated than \fg stars, influencing their present-day spatial and kinematic distributions.

We have shown that the observed overconcentration of \fg relative to \sg can be explained within the AGB scenario, provided certain internal and external conditions are met. This effect appears to be transient, driven by fluctuations in the number and spatial distribution of RGB stars. Further analysis is needed to confirm these results, particularly regarding binary fraction distributions and velocity anisotropy in \fg and \sg. In an upcoming study, we shall conduct a more detailed comparison of \mocca model results with observations from \citet{Leitingeretal2024} and \citet{Cadelano2024} to further investigate these trends.

\begin{acknowledgements}
We are grateful to the anonymous reviewer for their insightful comments, which helped improve the manuscript. MG, AH, GW, LH were supported by the Polish National Science Center (NCN) through the grant 2021/41/B/ST9/01191. AA acknowledges support for this paper from project No. 2021/43/P/ST9/03167 co-funded by the Polish National Science Center (NCN) and the European Union Framework Programme for Research and Innovation Horizon 2020 under the Marie Skłodowska-Curie grant agreement No. 945339. AH acknowledges support by the IDUB grant 140/04/POB4/0006 at Adam Mickiewicz University in Poznan, Poland.  
For the purpose of Open Access, the authors have applied for a CC-BY public copyright license to any Author Accepted Manuscript (AAM) version arising from this submission.
\end{acknowledgements} 


\section*{Data Availability} 
The data for the \mocca model described in this paper, including the 13 Gyr snapshot, is available at \url{https://doi.org/10.5281/zenodo.14908094}. Additional data can be provided upon request to the corresponding authors.


\bibliographystyle{aa}
\bibliography{ref.bib}

\begin{thebibliography}{54}
\expandafter\ifx\csname natexlab\endcsname\relax\def\natexlab#1{#1}\fi

\bibitem[{{Arca Sedda} {et~al.}(2018){Arca Sedda}, {Askar}, \&
  {Giersz}}]{ArcaSedda2018}
{Arca Sedda}, M., {Askar}, A., \& {Giersz}, M. 2018, \mnras, 479, 4652

\bibitem[{{Askar} {et~al.}(2018){Askar}, {Arca Sedda}, \&
  {Giersz}}]{Askar2018-bhsubsys}
{Askar}, A., {Arca Sedda}, M., \& {Giersz}, M. 2018, \mnras, 478, 1844

\bibitem[{{Baker} {et~al.}(2008){Baker}, {Boggs}, {Centrella}, {Kelly},
  {McWilliams}, {Miller}, \& {van Meter}}]{Bakeretal2008}
{Baker}, J.~G., {Boggs}, W.~D., {Centrella}, J., {et~al.} 2008, \apjl, 682, L29

\bibitem[{{Banerjee} {et~al.}(2020){Banerjee}, {Belczynski}, {Fryer},
  {Berczik}, {Hurley}, {Spurzem}, \& {Wang}}]{Banerjee2020}
{Banerjee}, S., {Belczynski}, K., {Fryer}, C.~L., {et~al.} 2020, \aap, 639, A41

\bibitem[{{Bastian} \& {Lardo}(2018)}]{Bastian2018}
{Bastian}, N. \& {Lardo}, C. 2018, \araa, 56, 83

\bibitem[{{Baumgardt} \& {Hilker}(2018)}]{Baumgardt2018}
{Baumgardt}, H. \& {Hilker}, M. 2018, \mnras, 478, 1520

\bibitem[{{Baumgardt} {et~al.}(2019){Baumgardt}, {Hilker}, {Sollima}, \&
  {Bellini}}]{Baumgardtetal2019}
{Baumgardt}, H., {Hilker}, M., {Sollima}, A., \& {Bellini}, A. 2019, \mnras,
  482, 5138

\bibitem[{{Belczynski} {et~al.}(2016){Belczynski}, {Heger}, {Gladysz},
  {Ruiter}, {Woosley}, {Wiktorowicz}, {Chen}, {Bulik}, {O'Shaughnessy}, {Holz},
  {Fryer}, \& {Berti}}]{Belczynskietal2016}
{Belczynski}, K., {Heger}, A., {Gladysz}, W., {et~al.} 2016, \aap, 594, A97

\bibitem[{{Belczynski} {et~al.}(2002){Belczynski}, {Kalogera}, \&
  {Bulik}}]{Belczynskietal2002}
{Belczynski}, K., {Kalogera}, V., \& {Bulik}, T. 2002, \apj, 572, 407

\bibitem[{{Belloni} {et~al.}(2017){Belloni}, {Askar}, {Giersz}, {Kroupa}, \&
  {Rocha-Pinto}}]{Bellonietal2017}
{Belloni}, D., {Askar}, A., {Giersz}, M., {Kroupa}, P., \& {Rocha-Pinto}, H.~J.
  2017, \mnras, 471, 2812

\bibitem[{{Belloni} {et~al.}(2018){Belloni}, {Kroupa}, {Rocha-Pinto}, \&
  {Giersz}}]{Bellonietal2018}
{Belloni}, D., {Kroupa}, P., {Rocha-Pinto}, H.~J., \& {Giersz}, M. 2018,
  \mnras, 474, 3740

\bibitem[{{Breen} \& {Heggie}(2013)}]{BreenHeggie2013}
{Breen}, P.~G. \& {Heggie}, D.~C. 2013, \mnras, 432, 2779

\bibitem[{{Cadelano} {et~al.}(2024){Cadelano}, {Dalessandro}, \&
  {Vesperini}}]{Cadelano2024}
{Cadelano}, M., {Dalessandro}, E., \& {Vesperini}, E. 2024, \aap, 685, A158

\bibitem[{{Chen} \& {Gnedin}(2024)}]{Chen2024}
{Chen}, Y. \& {Gnedin}, O.~Y. 2024, The Open Journal of Astrophysics, 7, 23

\bibitem[{{Dickson} {et~al.}(2024){Dickson}, {Smith}, {H{\'e}nault-Brunet},
  {Gieles}, \& {Baumgardt}}]{Dickson2024}
{Dickson}, N., {Smith}, P.~J., {H{\'e}nault-Brunet}, V., {Gieles}, M., \&
  {Baumgardt}, H. 2024, \mnras, 529, 331

\bibitem[{{Fregeau} {et~al.}(2004){Fregeau}, {Cheung}, {Portegies Zwart}, \&
  {Rasio}}]{Fregeauetal2004}
{Fregeau}, J.~M., {Cheung}, P., {Portegies Zwart}, S.~F., \& {Rasio}, F.~A.
  2004, \mnras, 352, 1

\bibitem[{{Fregeau} \& {Rasio}(2007)}]{Fregeau2007}
{Fregeau}, J.~M. \& {Rasio}, F.~A. 2007, \apj, 658, 1047

\bibitem[{{Fryer} {et~al.}(2012){Fryer}, {Belczynski}, {Wiktorowicz},
  {Dominik}, {Kalogera}, \& {Holz}}]{Fryeretal2012}
{Fryer}, C.~L., {Belczynski}, K., {Wiktorowicz}, G., {et~al.} 2012, \apj, 749,
  91

\bibitem[{{Fuller} \& {Ma}(2019)}]{Fuller2019}
{Fuller}, J. \& {Ma}, L. 2019, \apjl, 881, L1

\bibitem[{{Geisler} {et~al.}(1995){Geisler}, {Piatti}, {Claria}, \&
  {Minniti}}]{Geisleretal1995}
{Geisler}, D., {Piatti}, A.~E., {Claria}, J.~J., \& {Minniti}, D. 1995, \aj,
  109, 605

\bibitem[{{Giersz}(1998)}]{Giersz1998}
{Giersz}, M. 1998, \mnras, 298, 1239

\bibitem[{{Giersz} {et~al.}(2024){Giersz}, {Askar}, {Hypki}, {Hong},
  {Wiktorowicz}, \& {Hellstrom}}]{Gierszetal2024}
{Giersz}, M., {Askar}, A., {Hypki}, A., {et~al.} 2024, arXiv e-prints,
  arXiv:2411.06421

\bibitem[{{Giersz} {et~al.}(2013){Giersz}, {Heggie}, {Hurley}, \&
  {Hypki}}]{Gierszetal2013}
{Giersz}, M., {Heggie}, D.~C., {Hurley}, J.~R., \& {Hypki}, A. 2013, \mnras,
  431, 2184

\bibitem[{{Giesers} {et~al.}(2018){Giesers}, {Dreizler}, {Husser}, {Kamann},
  {Anglada Escud{\'e}}, {Brinchmann}, {Carollo}, {Roth}, {Weilbacher}, \&
  {Wisotzki}}]{Giesers2018}
{Giesers}, B., {Dreizler}, S., {Husser}, T.-O., {et~al.} 2018, \mnras, 475, L15

\bibitem[{{Giesers} {et~al.}(2019){Giesers}, {Kamann}, {Dreizler}, {Husser},
  {Askar}, {G{\"o}ttgens}, {Brinchmann}, {Latour}, {Weilbacher}, {Wendt}, \&
  et~al.}]{Giesers2019}
{Giesers}, B., {Kamann}, S., {Dreizler}, S., {et~al.} 2019, \aap, 632, A3

\bibitem[{{Gratton} {et~al.}(2019){Gratton}, {Bragaglia}, {Carretta},
  {D'Orazi}, {Lucatello}, \& {Sollima}}]{Gratton2019}
{Gratton}, R., {Bragaglia}, A., {Carretta}, E., {et~al.} 2019, \aapr, 27, 8

\bibitem[{{Harris}(1996)}]{Harris1996}
{Harris}, W.~E. 1996, \aj, 112, 1487

\bibitem[{{Hobbs} {et~al.}(2005){Hobbs}, {Lorimer}, {Lyne}, \&
  {Kramer}}]{Hobbsetal2005}
{Hobbs}, G., {Lorimer}, D.~R., {Lyne}, A.~G., \& {Kramer}, M. 2005, \mnras,
  360, 974

\bibitem[{{Hurley} {et~al.}(2000){Hurley}, {Pols}, \& {Tout}}]{Hurleyetal2000}
{Hurley}, J.~R., {Pols}, O.~R., \& {Tout}, C.~A. 2000, \mnras, 315, 543

\bibitem[{{Hurley} {et~al.}(2002){Hurley}, {Tout}, \& {Pols}}]{Hurleyetal2002}
{Hurley}, J.~R., {Tout}, C.~A., \& {Pols}, O.~R. 2002, \mnras, 329, 897

\bibitem[{{Hypki} \& {Giersz}(2013)}]{Hypki2013}
{Hypki}, A. \& {Giersz}, M. 2013, \mnras, 429, 1221

\bibitem[{{Hypki} {et~al.}(2022){Hypki}, {Giersz}, {Hong}, {Leveque}, {Askar},
  {Belloni}, \& {Otulakowska-Hypka}}]{Hypkietal2022}
{Hypki}, A., {Giersz}, M., {Hong}, J., {et~al.} 2022, \mnras, 517, 4768

\bibitem[{{Hypki} {et~al.}(2024){Hypki}, {Vesperini}, {Giersz}, {Hong},
  {Askar}, {Otulakowska-Hypka}, {Hellstrom}, \& {Wiktorowicz}}]{Hypkietal2024}
{Hypki}, A., {Vesperini}, E., {Giersz}, M., {et~al.} 2024, arXiv e-prints,
  arXiv:2406.08059

\bibitem[{{Kamlah} {et~al.}(2022){Kamlah}, {Leveque}, {Spurzem}, {Arca Sedda},
  {Askar}, {Banerjee}, {Berczik}, {Giersz}, {Hurley}, {Belloni},
  {K{\"u}hmichel}, \& {Wang}}]{Kamlahetal2022}
{Kamlah}, A.~W.~H., {Leveque}, A., {Spurzem}, R., {et~al.} 2022, \mnras, 511,
  4060

\bibitem[{{King}(1966)}]{King1966}
{King}, I.~R. 1966, \aj, 71, 64

\bibitem[{{Kremer} {et~al.}(2019){Kremer}, {Chatterjee}, {Ye}, {Rodriguez}, \&
  {Rasio}}]{Kremer2019}
{Kremer}, K., {Chatterjee}, S., {Ye}, C.~S., {Rodriguez}, C.~L., \& {Rasio},
  F.~A. 2019, \apj, 871, 38

\bibitem[{{Kremer} {et~al.}(2018){Kremer}, {Ye}, {Chatterjee}, {Rodriguez}, \&
  {Rasio}}]{Kremer2018}
{Kremer}, K., {Ye}, C.~S., {Chatterjee}, S., {Rodriguez}, C.~L., \& {Rasio},
  F.~A. 2018, \apjl, 855, L15

\bibitem[{{Kroupa}(2001)}]{Kroupa2001}
{Kroupa}, P. 2001, \mnras, 322, 231

\bibitem[{{Leitinger} {et~al.}(2024){Leitinger}, {Baumgardt}, {Cabrera-Ziri},
  {Hilker}, {Carbajo-Hijarrubia}, {Gieles}, {Husser}, \&
  {Kamann}}]{Leitingeretal2024}
{Leitinger}, E., {Baumgardt}, H., {Cabrera-Ziri}, I., {et~al.} 2024, arXiv
  e-prints, arXiv:2410.02855

\bibitem[{{Leitinger} {et~al.}(2023){Leitinger}, {Baumgardt}, {Cabrera-Ziri},
  {Hilker}, \& {Pancino}}]{Leitingeretal2023}
{Leitinger}, E., {Baumgardt}, H., {Cabrera-Ziri}, I., {Hilker}, M., \&
  {Pancino}, E. 2023, \mnras, 520, 1456

\bibitem[{{Mackey} {et~al.}(2008){Mackey}, {Wilkinson}, {Davies}, \&
  {Gilmore}}]{Mackey2008}
{Mackey}, A.~D., {Wilkinson}, M.~I., {Davies}, M.~B., \& {Gilmore}, G.~F. 2008,
  \mnras, 386, 65

\bibitem[{{Massari} {et~al.}(2019){Massari}, {Koppelman}, \&
  {Helmi}}]{Massarietal2019}
{Massari}, D., {Koppelman}, H.~H., \& {Helmi}, A. 2019, \aap, 630, L4

\bibitem[{{Mehta} {et~al.}(2024){Mehta}, {Milone}, {Casagrande}, {Marino},
  {Legnardi}, {Cordoni}, {Dondoglio}, {Jang}, {Lionetto}, {Ziliotto}, \&
  et~al.}]{Mehta2024}
{Mehta}, V.~J., {Milone}, A.~P., {Casagrande}, L., {et~al.} 2024, \mnras
  [\eprint[arXiv]{2406.02755}]

\bibitem[{{Milone} \& {Marino}(2022)}]{Milone2022}
{Milone}, A.~P. \& {Marino}, A.~F. 2022, Universe, 8, 359

\bibitem[{{Milone} {et~al.}(2017){Milone}, {Piotto}, {Renzini}, {Marino},
  {Bedin}, {Vesperini}, {D'Antona}, {Nardiello}, {Anderson}, {King}, {Yong},
  {Bellini}, {Aparicio}, {Barbuy}, {Brown}, {Cassisi}, {Ortolani}, {Salaris},
  {Sarajedini}, \& {van der Marel}}]{Miloneetal2017}
{Milone}, A.~P., {Piotto}, G., {Renzini}, A., {et~al.} 2017, \mnras, 464, 3636

\bibitem[{{Morawski} {et~al.}(2018){Morawski}, {Giersz}, {Askar}, \&
  {Belczynski}}]{Morawskietal2018}
{Morawski}, J., {Giersz}, M., {Askar}, A., \& {Belczynski}, K. 2018, \mnras,
  481, 2168

\bibitem[{{Morscher} {et~al.}(2015){Morscher}, {Pattabiraman}, {Rodriguez},
  {Rasio}, \& {Umbreit}}]{Morscher2015}
{Morscher}, M., {Pattabiraman}, B., {Rodriguez}, C., {Rasio}, F.~A., \&
  {Umbreit}, S. 2015, \apj, 800, 9

\bibitem[{{Peuten} {et~al.}(2016){Peuten}, {Zocchi}, {Gieles}, {Gualandris}, \&
  {H{\'e}nault-Brunet}}]{Peuten2016}
{Peuten}, M., {Zocchi}, A., {Gieles}, M., {Gualandris}, A., \&
  {H{\'e}nault-Brunet}, V. 2016, \mnras, 462, 2333

\bibitem[{{Renaud} {et~al.}(2017){Renaud}, {Agertz}, \&
  {Gieles}}]{Renaudetal2017}
{Renaud}, F., {Agertz}, O., \& {Gieles}, M. 2017, \mnras, 465, 3622

\bibitem[{{Sippel} \& {Hurley}(2013)}]{Sippel2013}
{Sippel}, A.~C. \& {Hurley}, J.~R. 2013, \mnras, 430, L30

\bibitem[{{Tanikawa} {et~al.}(2020){Tanikawa}, {Yoshida}, {Kinugawa},
  {Takahashi}, \& {Umeda}}]{Tanikawaetal2020}
{Tanikawa}, A., {Yoshida}, T., {Kinugawa}, T., {Takahashi}, K., \& {Umeda}, H.
  2020, \mnras, 495, 4170

\bibitem[{{Vitral} {et~al.}(2022){Vitral}, {Kremer}, {Libralato}, {Mamon}, \&
  {Bellini}}]{Vitral2022}
{Vitral}, E., {Kremer}, K., {Libralato}, M., {Mamon}, G.~A., \& {Bellini}, A.
  2022, \mnras, 514, 806

\bibitem[{{Weatherford} {et~al.}(2020){Weatherford}, {Chatterjee}, {Kremer}, \&
  {Rasio}}]{Weatherford2020}
{Weatherford}, N.~C., {Chatterjee}, S., {Kremer}, K., \& {Rasio}, F.~A. 2020,
  \apj, 898, 162

\bibitem[{{Zocchi} {et~al.}(2019){Zocchi}, {Gieles}, \&
  {H{\'e}nault-Brunet}}]{Zocchi2019}
{Zocchi}, A., {Gieles}, M., \& {H{\'e}nault-Brunet}, V. 2019, \mnras, 482, 4713

\end{thebibliography}

\begin{appendix}

\section{MOCCA code}
\label{s:Appendix0}

This work is based on the numerical simulations performed with the \mocca Monte Carlo code \citep[][]{Giersz1998, Hypki2013, Gierszetal2013, Hypkietal2022, Hypkietal2024, Gierszetal2024}. \mocca is an advanced code that performs full stellar and dynamical evolution of real-size star clusters up to the Hubble time. The implementation of stellar and binary evolution within the \mocca code is based on the rapid population synthesis code \bse code \citep[][]{Hurleyetal2000, Hurleyetal2002} that has been strongly updated by \citet[][]{Bellonietal2017, Bellonietal2018}, \citet[][]{Banerjee2020}, and \citet[][]{Kamlahetal2022}.
Strong dynamical interactions in \mocca are performed with \fewbody code \citep[][]{Fregeauetal2004, Fregeau2007}.
\mocca can follow the full dynamical and stellar evolution of \mpop, allowing for delayed \sg formation relative to \fg. In this scenario, \sg forms through the re-accretion of gas surrounding the newly formed cluster, which mixes with enriched material from the ejected envelopes of AGB stars. Additionally, \mocca incorporates the effects of cluster migration, driven by the rapidly evolving gravitational potential of the early Galaxy and dynamical friction.

The stellar and binary evolution parameters used in the presented simulations are referred to as Model C in \citet{Kamlahetal2022}. In short, the metallicity of both populations in all the simulated models was set to Z = 0.001. The updated treatment for the evolution of massive stars was used according to \citet{Tanikawaetal2020} together with improved treatment for mass loss due to stellar winds and the inclusion of pair and pulsational pair-instability supernova \citep[][]{Belczynskietal2016}. The masses of BHs and NS were determined according to the rapid supernovae prescriptions from \citet{Fryeretal2012}. NS natal kicks were sampled from a Maxwelllian distribution with $\sigma = 265$ km/s \citep{Hobbsetal2005}. However, for BHs, these natal kicks were reduced according to the mass fallback prescription \citep[][]{Belczynskietal2002, Fryeretal2012}. The formation of neutron stars with negligible natal kicks through electron-capture or accretion-induced supernova was also enabled. Another feature of these models is the inclusion of gravitational wave recoil kicks whenever two BHs merge \citep[][]{Bakeretal2008, Morawskietal2018}. Low birth spins were assumed for BHs, with values uniformly sampled between 0 and 0.1 \citep[][]{Fuller2019}. The orientation of the BH spin relative to the binary orbit was randomly distributed \citep{Morawskietal2018}.
The full list of changes in the recent \mocca code, together with their detailed description, can be found in \citet{Gierszetal2024}.

\section{BH subsystem evolution}\label{s:Appendix1}

\begin{figure}
\begin{center}
  \includegraphics[width=0.99\linewidth]{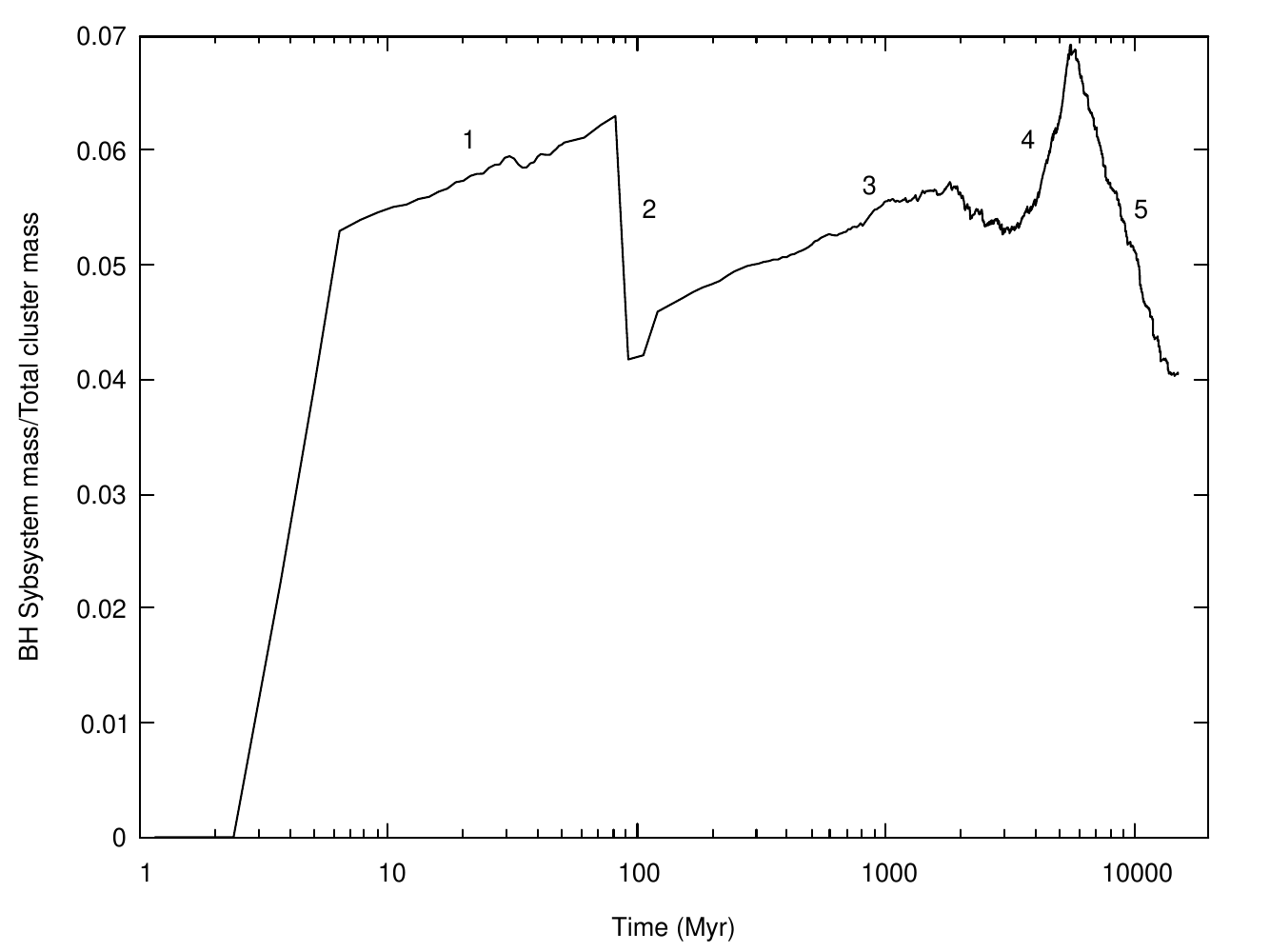}
  \caption{Time evolution of the BH subsystem mass as a fraction of the total cluster mass. Labels 1–5 mark the distinct evolutionary phases described in Section \ref{s:Appendix1}}. 
  \label{f:A1}
\end{center}
\end{figure}

The BHS plays a critical role in the development of 1P overconcentration and evolves differently from that in single-population GCs that are initially tidally underfilling. Figure\,\ref{f:A1} shows the time evolution of the BHS-to-cluster mass ratio, highlighting five key phases. 

\begin{figure}
\begin{center}
  \includegraphics[width=0.99\linewidth]{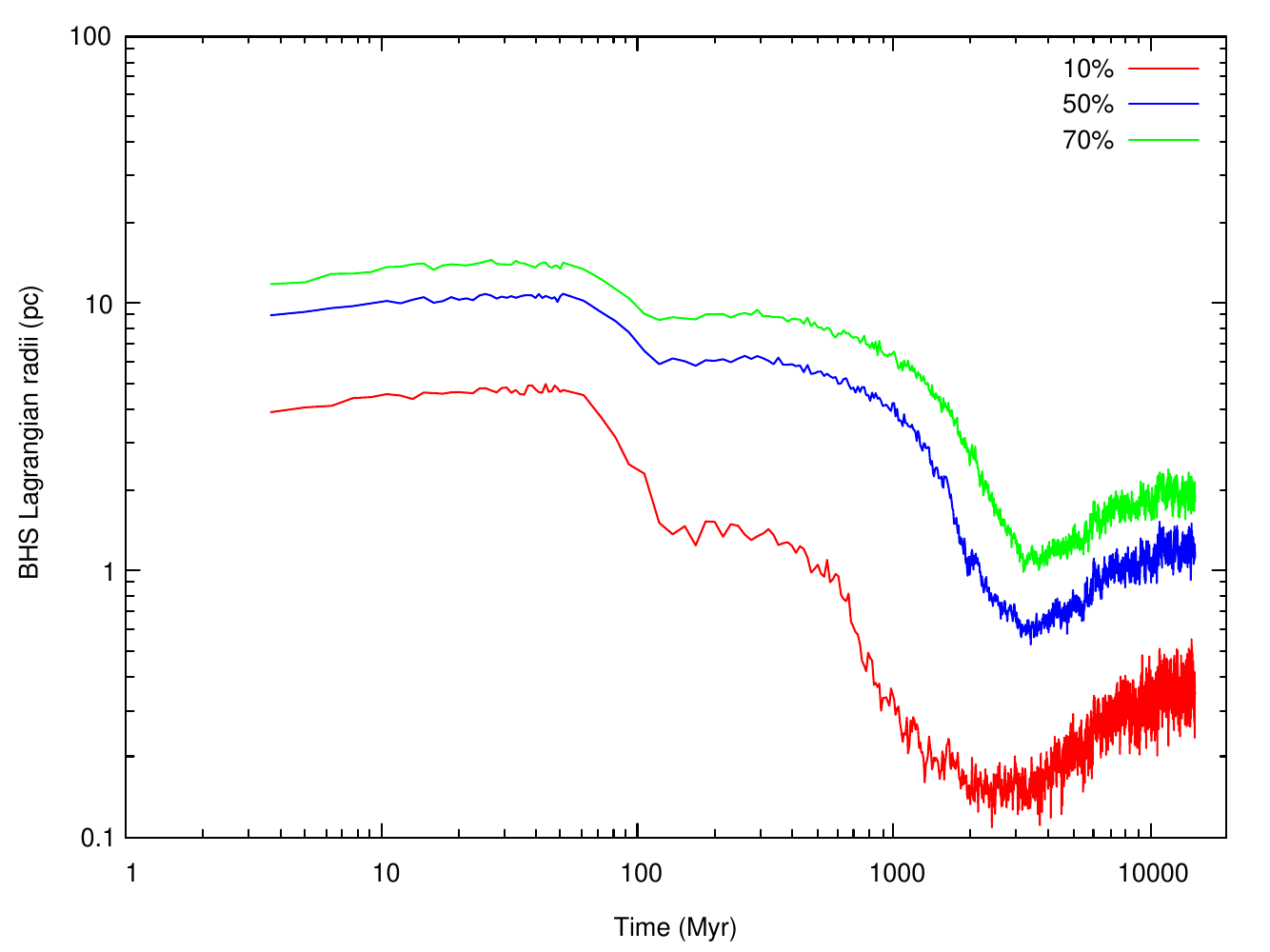}
  \caption{Time evolution of Lagrangian radii (10\%, 50\%, and 70\%) for BHs, showing the progressive central concentration of the BHS}
  \label{f:A2}
\end{center}
\end{figure}

In phase 1, the BHS mass fraction rises rapidly due to BH formation from massive 1P stars and concurrent 1P mass loss as the cluster expands and overfills its tidal radius. In phase 2, the mass fraction drops following 2P formation, which significantly increases the total mass while contributing few BHs. The evolution beyond this point is shaped by BHS mass segregation and contraction (see Fig.\,\ref{f:A2}). In phase 3, this segregation fuels energy generation that drives enhanced stellar escape, causing the BHS mass fraction to rise again. The brief dip around 2 Gyr reflects the final stages of BHS formation and early BH ejections. Phase 4 marks a period of balanced evolution, with the cluster expanding and preferentially losing 1P stars. As the system is only mildly nTF, escape remains efficient, and the BHS mass fraction increases further. At 5.5 Gyr, the cluster migrates to a larger Galactocentric distance and becomes strongly nTF (phase 5), suppressing stellar escape and leading to gradual BHS depletion through internal interactions. While the present-day BHS mass fraction in our model (5–6\%) is higher than values inferred for most Galactic GCs—which are typically below 1\% \citep[][]{Weatherford2020,Dickson2024}—it is comparable to estimates for exceptional cases such as Omega Centauri \citep{Zocchi2019,Baumgardtetal2019}. For clusters like NGC 3201 and NGC 6101, the inferred BH mass fractions are typically lower, around 1–2\%. However, we note that multi-mass modeling techniques used to derive these estimates can systematically underpredict BH mass fractions above ~1\%, particularly in dynamically young clusters \citep[see Fig. 1 in][]{Dickson2024}. Although NGC~3201, similar to our model, has a large present-day half-mass radius and appears dynamically young, our simulation suggests that it underwent significant dynamical evolution prior to its migration to a larger Galactocentric distance.

\section{Observational core radius}\label{s:Appendix2}

\begin{figure}[h]
\begin{center}
  \includegraphics[width=0.99\linewidth]{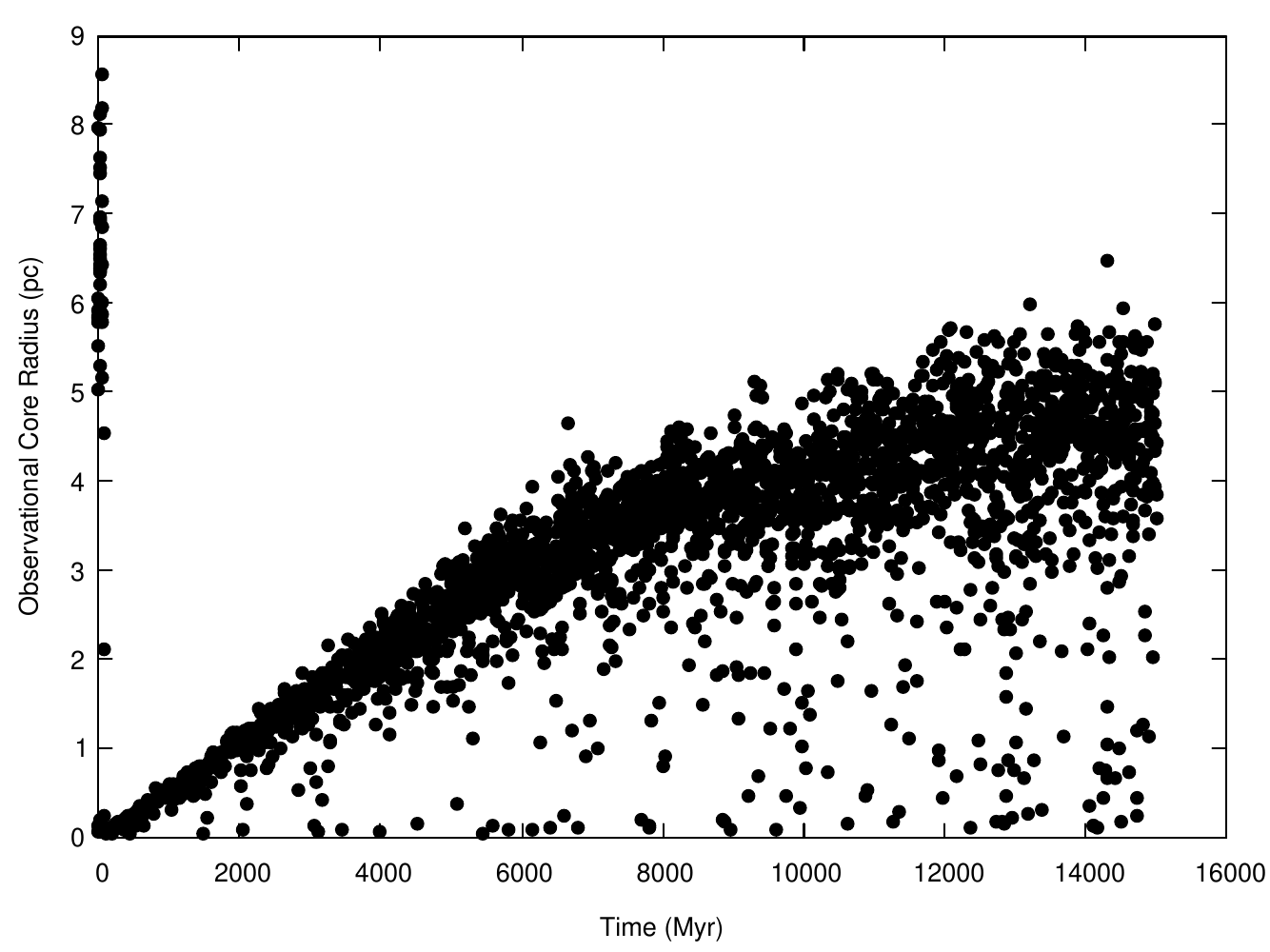}
  \caption{Observational core radius as a function of time.}
  \label{f:A3}
\end{center}
\end{figure}

The observational core radius, as calculated in MOCCA, can vary substantially between output times, particularly at late stages of cluster evolution. This variability arises from the small number of bright evolved stars that dominate the central surface brightness, making the derived core radius sensitive to their spatial distribution. 
Figure\,\ref{f:A3} shows the time evolution of the observational core radius in our model. While the long-term trend reflects the global dynamical state of the cluster, significant variations between output times are apparent --ranging from below 1 pc to over 6 pc. These fluctuations are not necessarily indicative of structural changes, but rather reflect changes in the central surface brightness profile caused by the positions of a few luminous stars.
For example, at the output time shown in the main paper (13.002 Gyr), the observational core radius is 3.13 pc and the central surface brightness (CSB) is 610\,L$_\odot$/pc$^2$. In the immediately following output (13.006 Gyr), these values shift to 2.07 pc and 817\,L$_\odot$/pc$^2$, respectively—closer to the observational values reported for NGC 3201 in the \citet[][updated 2010]{Harris1996} catalog, which lists a core radius of 1.85 pc and a CSB of 914\,L$_\odot$/pc$^2$.
Given these fluctuations, we emphasize that other structural parameters—such as total luminosity, cluster mass, and half-light radius—are more stable and provide a more reliable basis for comparison with observations. Additionally, the 3D core radius of the model at 13.002 Gyr is 1.22 pc, with a half-mass radius of 8.71 pc. At 13.006 Gyr, the 3D core radius is 1.77 pc, and the half-mass radius is 8.72 pc. These values are in excellent agreement with the parameters listed for NGC 3201 in the \citet[][updated 2024]{Baumgardt2018} catalog\footnote{\url{https://people.smp.uq.edu.au/HolgerBaumgardt/globular/parameter.html}}.

\end{appendix}

\end{document}